\documentstyle[emulateapj]{article}

\lefthead{Nagao, Murayama, \& Taniguchi}
\righthead{[O {\sc iii}]$\lambda$4363 Emitting Region}

\begin{document}

\submitted{The Astrophysical Journal, 549, in press}
\title{WHERE IS THE [O {\sc iii}]$\lambda$4363 EMITTING REGION \\
       IN ACTIVE GALACTIC NUCLEI?}

\author{Tohru NAGAO$^1$, Takashi MURAYAMA$^1$, and Yoshiaki TANIGUCHI$^{1,2}$}
\affil{$^1$ Astronomical Institute, Graduate School of Science, 
       Tohoku University, Aramaki, Aoba, Sendai 980-8578, Japan}
\affil{$^2$ Institute for Astronomy, University of Hawaii,
       2680 Woodlawn Drive, Honolulu, HI 96822}

\begin{abstract}

The emission-line flux ratio of 
[O {\sc iii}]$\lambda$4363/[O {\sc iii}]$\lambda$5007 ($R_{\rm O III}$)
is a useful diagnostic for the ionization mechanism 
and physical properties of emission-line regions in active galactic
nuclei (AGNs). 
However, it is known that simple photoionization models underpredict
the [O {\sc iii}]$\lambda$4363 intensity, being inconsistent with 
observations. 
In this paper, we report on several pieces of evidence that
a large fraction of the [O {\sc iii}]$\lambda$4363 emission arises 
from the dense gas obscured by putative tori:
(1) the visibility of high-$R_{\rm O III}$ regions is
correlated to that of broad-line regions, (2) higher-$R_{\rm O III}$
objects show hotter mid-infrared colors, (3) higher-$R_{\rm O III}$ 
objects show stronger highly-ionized emission lines such as 
[Fe {\sc vii}]$\lambda$6087 and [Fe {\sc x}]$\lambda$6374, and 
(4) higher-$R_{\rm O III}$ objects have broader line width of
[O {\sc iii}]$\lambda$4363
normalized by that of [O {\sc iii}]$\lambda$5007. 
To estimate how such a dense
component contributes to the total emission-line flux, dual-component 
photoionization model calculations are performed.
It is shown that the observed values of $R_{\rm O III}$ of type 1 AGNs 
may be explained by introducing 
a 5\% -- 20\% contribution from the dense component while 
those of type 2 AGNs may be explained by introducing 
a 0\% -- 2\% contribution. 
We also discuss the [O {\sc iii}]$\lambda$4363 emitting regions in LINERs
in the framework of our dual-component model.

\end{abstract}


\section{INTRODUCTION}

It has been often considered that emission-line regions around active galactic
nuclei (AGNs) are photoionized by the nonthermal continuum radiation from 
central engines (e.g., Davidson 1977; Yee 1980; Kwan \& Krolik 1981; 
Shuder 1981; Cohen 1983; Cruz-Gonz\'alez et al. 1991; Osterbrock 1993; 
Evans et al. 1999).
However, this photoionization scenario has sometimes been confronted with 
several serious problems. 

One of such problems is that 
any single-zone photoionization models underpredict 
the [O {\sc iii}]$\lambda$4363/[O {\sc iii}]$\lambda$5007 intensity ratio,
$R_{\rm O III}$ (e.g., Koski \& Osterbrock 1976; Ferland \& Netzer 1983; 
Filippenko \& Halpern 1984; Viegas-Aldrovandi \& Gruenwald 1988). 
The reason for the underprediction of $R_{\rm O III}$ is thought to be that
photoionization of gas in optically thick condition is hard to accomplish
electron temperatures above a few $\times 10^4$ K
if density of the gas is typical in narrow-line regions (NLRs).
In order to solve this problem, many studies have been carried out.
Such attempts can be roughly divided into the following two categories.
One is based on the idea that a high density component may contribute to 
achieve the observed high $R_{\rm O III}$ (e.g., Baldwin 1975; 
Osterbrock, Koski, \& Phillips 1976; Filippenko \& Halpern 1984; 
Filippenko 1985).
This idea is attributed to the fact that
the critical density ($n_{\rm cr}$) of the [O {\sc iii}]$\lambda$4363 emission 
(3.3 $\times 10^7$ cm$^{-3}$) is higher than that of the 
[O {\sc iii}]$\lambda$5007 emission (7.0 $\times 10^5$ cm$^{-3}$);
which leads to high $R_{\rm O III}$ when the gas density is higher than
$\sim 10^6$ cm$^{-3}$.
The other idea is to introduce high temperature regions whose temperature
is more than a few times 10$^4$ K.
To achieve such high temperatures, either shock-heated regions 
(e.g., Koski \& Osterbrock 1976; Heckman 1980; Dopita \& Sutherland 1995) or
optically-thin components (e.g., Wilson, Binette, \& Storchi-Bergmann 1997) 
is required.

In addition, some previous studies reported that the values of
$R_{\rm O III}$ depend on the AGN type; i.e., type 1 Seyfert nuclei (S1s) 
exhibit higher $R_{\rm O III}$ than type 2 Seyfert nuclei (S2s) 
(e.g., Osterbrock et al. 1976;
Heckman \& Balick 1979; Shuder \& Osterbrock 1981;
Cohen 1983). 
This tendency seems to suggest that a part of
the [O {\sc iii}]$\lambda$4363 emission arises from 
the regions which are obscured only in S2s by any materials,
such as dusty tori.
However, this tendency may be interpreted by the intrinsic 
(that is, not due to 
obscuration effects) difference of NLR properties, such as the size 
(Schmitt \& Kinney 1996; Kraemer et al. 1998) or the ionization states 
(Schmitt 1998).

To investigate the reason why S1s show high $R_{\rm O III}$ than S2s
seems useful to explore where and how
the [O {\sc iii}]$\lambda$4363 emission is radiated.
Moreover, these may lead to the new solution to the underprediction
problem of $R_{\rm O III}$. Therefore,
in this paper, we present how the observed values of $R_{\rm O III}$ are
different among various types of Seyferts based on a large sample 
of Seyferts compiled from the literature.
Then, we compare $R_{\rm O III}$ with various parameters and discuss the
nature of the [O {\sc iii}]$\lambda$4363 emitting regions in AGNs.

\section{DATA COMPILATION}

 \subsection{Data}

We briefly summarize the policy of classification of Seyfert nuclei
in this paper, which is the same as that 
in  Nagao, Taniguchi, \& Murayama (2000c).
Seyfert nuclei are often divided into three types based on the visibility
of broad components of hydrogen recombination lines: 
i.e., S1, Seyfert 1.5 (S1.5), and S2.
The S1s consist of typical S1s (BLS1s; broad-line Seyfert 1 galaxies) and
narrow-line Seyfert 1 galaxies (NLS1s; e.g., Osterbrock \& Pogge 1985;
Boller, Brandt, \& Fink 1996). 
The type of Seyfert 1.2 (S1.2) is included in the type of BLS1 in this paper.
We divide the type of S2 into S2$^+$ and S2$^-$; the former one exhibits
the evidence for the existence of broad-line regions (BLRs) and the latter
does not show such the evidence.
For the convenience, both types of S2$^+$ and S2$^-$ are referred as 
S2$_{\rm total}$ when needed; i.e., S2$_{\rm total}$ = S2$^+$ + S2$^-$.
There are two populations in the types of S2$^+$: one shows weak symptoms
of the existence of BLRs in their optical or near-infrared (NIR) spectra
(S2$_{\rm RBLR}$; type 2 Seyfert with the reddened BLR) and
another one exhibits the hidden BLR which is detected only
in polarized spectra (S2$_{\rm HBLR}$).
The type of S2$_{\rm RBLR}$ consists of Seyfert 1.8 galaxies (S1.8s),
Seyfert 1.9 galaxies (S1.9s), and the Seyfert galaxies with the broad
Paschen or Bracket line (S2$_{\rm NIR-BLR}$s).
They are combined into S2$^+$ when statistical treatments are needed.

In order to investigate statistical properties of the observed values of
$R_{\rm O III}$ for each type of Seyferts, we compiled $R_{\rm O III}$
from the literature. The number of compiled objects is 214;
26 NLS1s, 56 BLS1s, 54 S1.5s, 4 S1.8s, 16 S1.9s, 5 S2$_{\rm NIR-BLR}$s,
8 S2$_{\rm HBLR}$s, and 45 S2$^-$s.
We basically referred to V\'{e}ron-Cetty \& V\'{e}ron (1998) for
the AGN type of each object. Though
Mrk 335, Mrk 766, Mrk 1126, H 34.06, H 1934--063, HE 1029--1831, and
J 13.12 are not classified as NLS1 by V\'{e}ron-Cetty \& V\'{e}ron (1998),
they are treated as NLS1 in this paper because 
they have been classified as NLS1 (Osterbrock \& Pogge 1985;
Vaughan et al. 1999; Rodr\'{\i}guez-Ardia, Pastoriza, \& Donzelli 2000).

All the objects are listed in Table 1 together with their redshifts,
$R_{\rm O III}$, and the fluxes at 3.5$\mu$m, 12$\mu$m, 25$\mu$m, and 60$\mu$m.
The values of $R_{\rm O III}$ in this table are the averaged ones among the
references given there.
We do not make the reddening correction for the values of $R_{\rm O III}$ since
it is often difficult to measure the narrow Balmer component, particularly
for S1s. 
It is noted that we do not use any upper limit data in this study.

 \subsection{Selection Bias}

\begin{figure*}
\epsscale{0.45}
\plotone{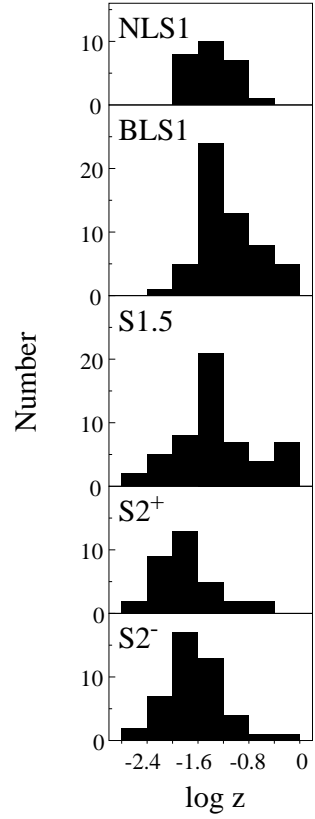}
\caption{
The frequency distributions of the redshift for the NLS1s, the BLS1s,
the S1.5s, the S2$^+$s, and the S2$^-$s.
\label{fig1}}
\end{figure*}

\begin{figure*}
\epsscale{1.0}
\plotone{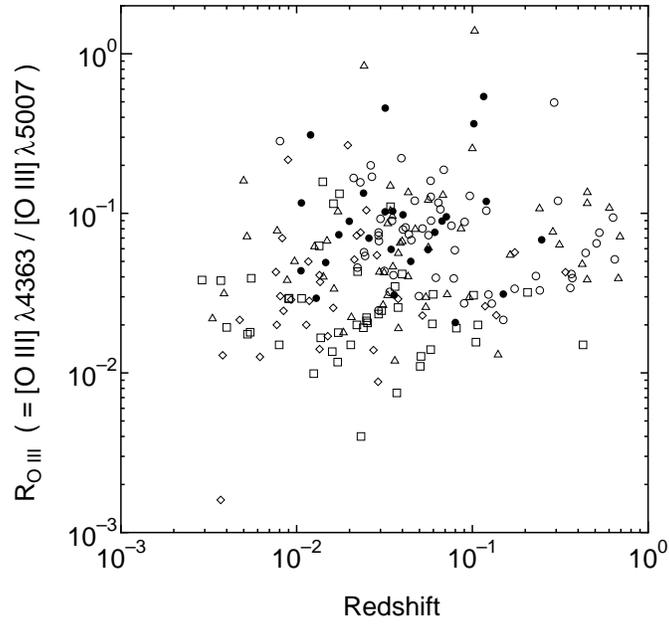}
\caption{
$R_{\rm O III}$ are plotted as a function of the redshift.
The NLS1s, the BLS1s,
the S1.5s, the S2$^+$s, and the S2$^-$s are
shown by filled circles, open ones, triangles, diamonds, and squares, 
respectively.
\label{fig2}}
\end{figure*}

\begin{figure*}
\epsscale{0.65}
\plotone{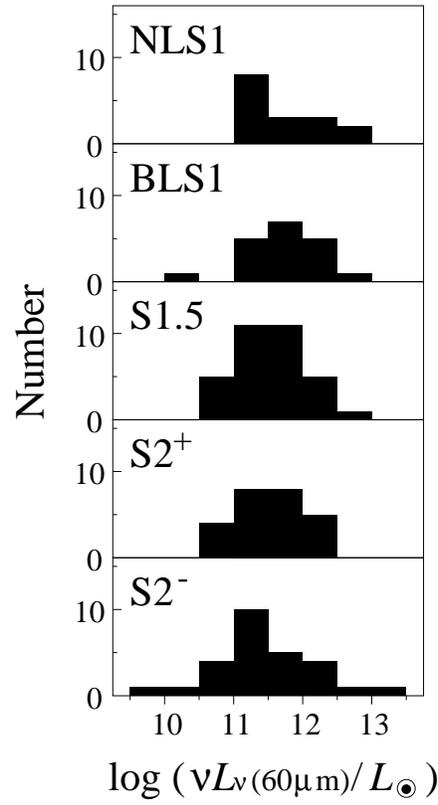}
\caption{
The frequency distributions of the 60$\mu$m luminosity 
for the NLS1s, the BLS1s,
the S1.5s, the S2$^+$s, and the S2$^-$s.
The luminosities are normalized by the solar luminosity.
\label{fig3}}
\end{figure*}

\begin{figure*}
\epsscale{1.0}
\plotone{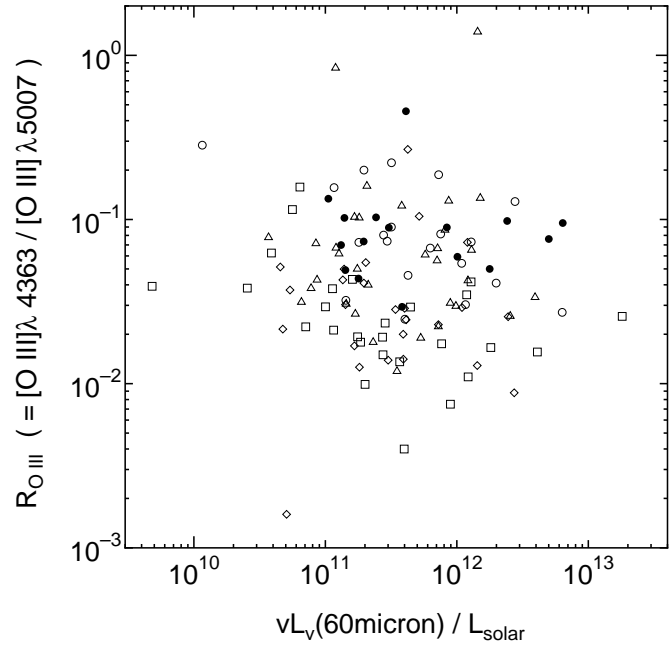}
\caption{
$R_{\rm O III}$ are plotted as a function of the 60$\mu$m luminosity.
The symbols are the same as in Figure 2.
\label{fig4}}
\end{figure*}

\begin{deluxetable}{lccc}
\tablenum{2}
\tablecaption{Means and Deviations of Redshift, $\nu L_{\nu}$(60$\mu$),
              and $R_{\rm O III}$ for Each Types of Seyfert Galaxies\label{tbl-2}}
\tablewidth{0pt}
\tablehead{
\colhead{type\tablenotemark{a}} &
\colhead{Redshift} &
\colhead{log ($\nu L_{\nu}$(60$\mu$m)/$L_{\rm \odot}$)} &
\colhead{$R_{\rm O III}$}
}
\startdata
NLS1 (26)                                 & 0.0567 $\pm$ 0.0532 & 11.694 $\pm$ 0.563 & 0.126 $\pm$ 0.132 \nl
BLS1 (56)                                 & 0.1329 $\pm$ 0.1612 & 11.664 $\pm$ 0.571 & 0.091 $\pm$ 0.077 \nl
S1.5 (54)                                 & 0.1192 $\pm$ 0.1768 & 11.546 $\pm$ 0.499 & 0.107 $\pm$ 0.210 \nl
S2$_{\rm total}$ (78)                     & 0.0407 $\pm$ 0.0674 & 11.454 $\pm$ 0.621 & 0.039 $\pm$ 0.044 \nl
\ \ \ \ \ S2$^+$ (33)                     & 0.0375 $\pm$ 0.0657 & 11.480 $\pm$ 0.496 & 0.047 $\pm$ 0.054 \nl
\ \ \ \ \ \ \ \ \ \ S1.8 (4)              & 0.0996 $\pm$ 0.1383 & 11.694 $\pm$ 0.387 & 0.097 $\pm$ 0.070 \nl
\ \ \ \ \ \ \ \ \ \ S1.9 (16)             & 0.0255 $\pm$ 0.0401 & 11.368 $\pm$ 0.495 & 0.053 $\pm$ 0.060 \nl
\ \ \ \ \ \ \ \ \ \ S2$_{\rm NIR-BLR}$ (5)& 0.0594 $\pm$ 0.0564 & 11.368 $\pm$ 0.109 & 0.029 $\pm$ 0.021 \nl
\ \ \ \ \ \ \ \ \ \ S2$_{\rm HBLR}$ (8)   & 0.0168 $\pm$ 0.0104 & 11.635 $\pm$ 0.522 & 0.023 $\pm$ 0.011 \nl
\ \ \ \ \ S2$^-$ (45)                     & 0.0431 $\pm$ 0.0685 & 11.429 $\pm$ 0.717 & 0.033 $\pm$ 0.032 \\
\enddata
\tablenotetext{a}{The number of objects for each type is written in parenthesis.}
\end{deluxetable}

\begin{deluxetable}{lccccc}
\tablenum{3}
\tablecaption{The Results of the KS Test 
              Concerning the Redshift, $\nu L_{\nu}$(60$\mu$),
              and $R_{\rm O III}$. \label{tbl-7}}
\tablewidth{0pt}
\tablehead{
\colhead{Type} &
\colhead{NLS1} &
\colhead{BLS1} & 
\colhead{S1.5} & 
\colhead{S2$^+$} &
\colhead{S2$^-$}
}
\startdata
\cutinhead{Redshift}
NLS1   & \nodata & 1.671 $\times 10^{-1}$ & 6.734 $\times 10^{-1}$ & 1.831 $\times 10^{-3}$ & 8.579 $\times 10^{-2}$ \nl
BLS1   & \nodata & \nodata                & 6.818 $\times 10^{-2}$ & 3.525 $\times 10^{-9}$ & 5.410 $\times 10^{-6}$ \nl
S1.5   & \nodata & \nodata                & \nodata                & 5.872 $\times 10^{-6}$ & 3.685 $\times 10^{-3}$ \nl
S2$^+$ & \nodata & \nodata                & \nodata                & \nodata                & 1.624 $\times 10^{-1}$ \nl
S2$^-$ & \nodata & \nodata                & \nodata                & \nodata                & \nodata                \nl
\cutinhead{$\nu L_{\nu}$(60$\mu$m)}
NLS1   & \nodata & 8.792 $\times 10^{-1}$ & 7.908 $\times 10^{-1}$ & 8.970 $\times 10^{-1}$ & 3.923 $\times 10^{-1}$ \nl
BLS1   & \nodata & \nodata                & 4.439 $\times 10^{-1}$ & 5.986 $\times 10^{-1}$ & 1.920 $\times 10^{-1}$ \nl
S1.5   & \nodata & \nodata                & \nodata                & 4.788 $\times 10^{-1}$ & 6.648 $\times 10^{-1}$ \nl
S2$^+$ & \nodata & \nodata                & \nodata                & \nodata                & 7.911 $\times 10^{-1}$ \nl
S2$^-$ & \nodata & \nodata                & \nodata                & \nodata                & \nodata                \nl
\cutinhead{$R_{\rm O III}$}
NLS1   & \nodata & 7.877 $\times 10^{-1}$ & 2.037 $\times 10^{-1}$ & 1.555 $\times 10^{-4}$ & 9.933 $\times 10^{-9}$ \nl
BLS1   & \nodata & \nodata                & 4.064 $\times 10^{-1}$ & 5.735 $\times 10^{-5}$ & 4.612 $\times 10^{-10}$ \nl
S1.5   & \nodata & \nodata                & \nodata                & 2.045 $\times 10^{-3}$ & 4.106 $\times 10^{-6}$ \nl
S2$^+$ & \nodata & \nodata                & \nodata                & \nodata                & 1.264 $\times 10^{-1}$ \nl
S2$^-$ & \nodata & \nodata                & \nodata                & \nodata                & \nodata                \nl
\enddata
\end{deluxetable}

Because we do not impose any selection criteria upon our sample, it is 
necessary to test whether or not the various samples are appropriate
for our comparative study. There would be possible biases if there are any
systematic differences in the redshift distributions or in the intrinsic 
nuclear luminosity distributions, 
thus we investigate those distributions below.

First we investigate the redshift distributions. We show the histograms
of redshift in Figure 1. 
The mean redshift and the 1$\sigma$ deviation for each type
are given in Table 2. 
There seems to be a tendency that the redshifts of the objects in the samples
of the BLS1 and the S1.5 are larger than those in the other samples.
In order to confirm whether or not this tendency is statistically real,
we apply the Kolmogorov-Smirnov (KS) statistical test (see Press et al. 1988).
The null hypothesis is that the redshift distributions among the NLS1s,
the BLS1s, the S1.5s, the S2$^+$s, and the S2$^-$s come from the same
underlying population. The KS probabilities are given in Table 3.
The KS test leads to the following results.
(1) The redsifts of the NLS1s, of the BLS1s, and of the S1.5s are
statistically indistinguishable. (2) The redshifts of the S2$^+$ and
of the S2$^-$ are also statistically indistinguishable. (3) However,
the former and the latter are statistically different.
Does this difference of the redshift cause any possible biases against the 
following comparative study? To investigate this issue, we examine
the relation between $R_{\rm O III}$, 
which is our main interest in this paper, 
and redshift (Figure 2). Figure 2 suggests that there is no correlation
between $R_{\rm O III}$ and redshift. This means that the redshift difference
among the samples is thought not to cause a bias against the investigation
of properties of $R_{\rm O III}$.

Second we consider the intrinsic AGN power.
The so-called AGN unified model (Antonucci \& Miller 1985; see for a review 
Antonucci 1993) requires anisotropic nuclear radiation.
This may cause systematic differences in intrinsic AGN power
among the types of Seyferts depending on selection criteria.
Statistical properties of emission-line ratios for each type of Seyferts
might suffer from this bias of intrinsic luminosity. Therefore,
we investigate the intrinsic AGN power distributions using the
{\it IRAS} 60$\mu$m luminosity, which is regarded as rather isotropic emission
(e.g., Pier \& Krolik 1992; Efstathiou \& Rowan-Robinson 1995; 
Fadda et al. 1998)
though this might be contaminated with the influence of circumnuclear 
star formation.
The histograms of the 60$\mu$m luminosity are shown in Figure 3.
The mean 60$\mu$m luminosities and 1$\sigma$ deviations are given in Table 2.
Here we adopt a Hubble constant $H_{\rm 0}$ = 50 km s$^{-1}$
Mpc$^{-1}$ and a deceleration parameter $q_{\rm 0}$ = 0.
We apply the KS test where the null hypothesis is that the distribution of
the 60$\mu$m luminosity of the samples come from the same underlying 
population. The resultant KS probabilities are given in Table 3.
The KS test suggests that there is no systematic difference 
in the 60$\mu$m luminosity among the types of Seyfert galaxies.
It is noted that there is no correlation between $R_{\rm O III}$ and 
the 60$\mu$m luminosity, i.e., the intrinsic AGN power (Figure 4).

\section{RESULTS}

 \subsection{Dependence of $R_{\rm O III}$ on the AGN Type}

\begin{figure*}
\epsscale{0.75}
\plotone{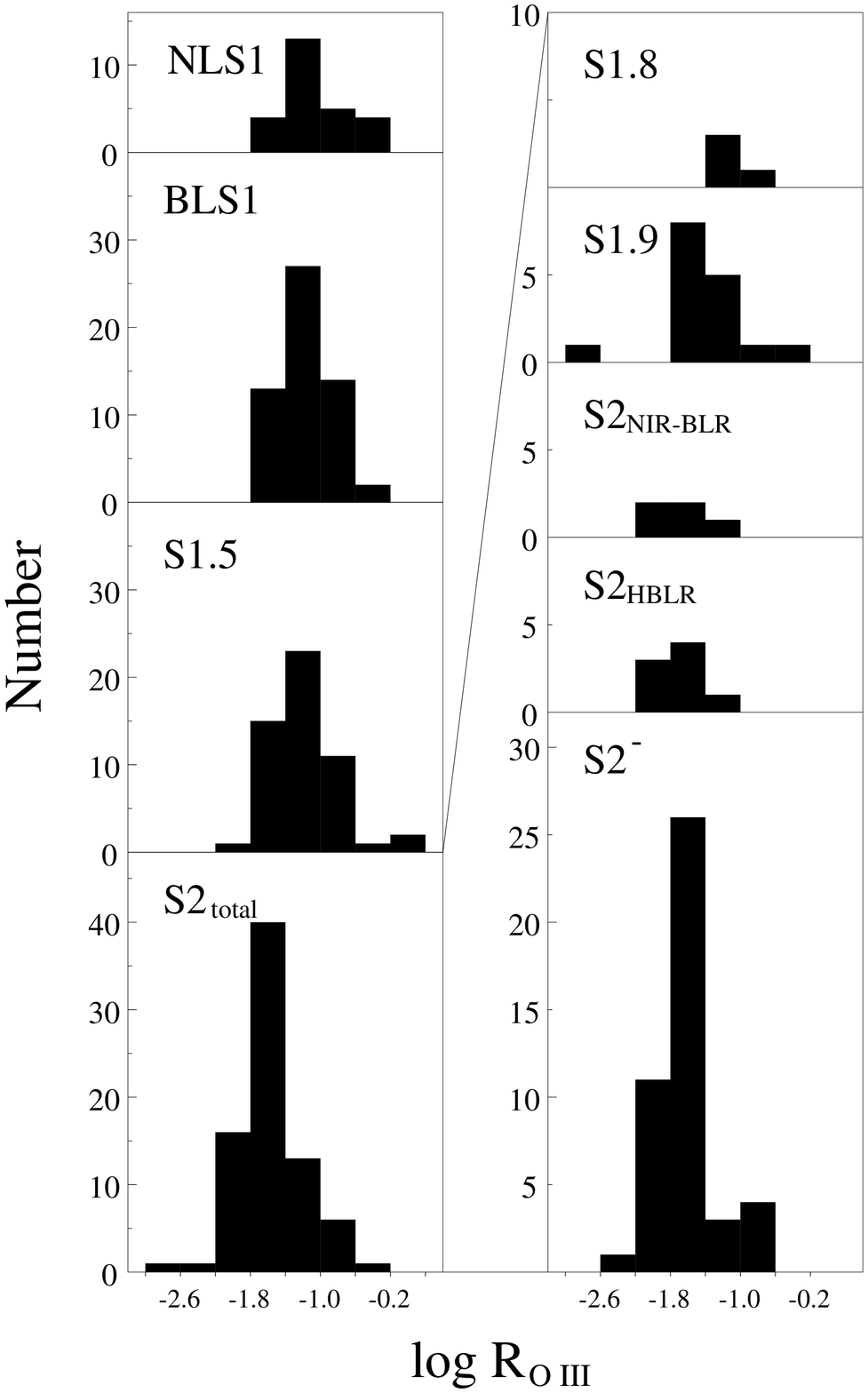}
\caption{
The frequency distributions of $R_{\rm O III}$ for the NLS1s, the BLS1s,
the S1.5s, and the S2$_{\rm total}$s.
The details of the frequency distributions of the S2$_{\rm total}$s are
also shown: i.e., for the S1.8s, the S1.9s, the S2$_{\rm NIR-BLR}$s,
the S2$_{\rm HBLR}$s, and the S2$^-$s.
\label{fig5}}
\end{figure*}

We show the histograms of $R_{\rm O III}$ for each type of Seyfert galaxies
in Figure 5. 
The mean and the 1$\sigma$ deviation of $R_{\rm O III}$ 
for each type are given in Table 2. 
In Figure 5, it is clearly shown that the S2$_{\rm total}$s exhibit
lower $R_{\rm O III}$ than the NLS1s, the BLS1s, and 
the S1.5s. There seems to be a tendency also in details of 
the S2$_{\rm total}$s; i.e., the values of $R_{\rm O III}$ of the S1.8s
are higher than those of the S1.9s, and those of the S1.9s are
higher than those of the S2$_{\rm NIR-BLR}$s, the S2$_{\rm HBLR}$s, and 
the S2$^-$s although the numbers of the samples are small.
These properties can be interpreted that
the more the BLR emission suffers the reddening, 
the lower the observed value of 
$R_{\rm O III}$ is.
We apply the KS test where the null hypothesis is that the distribution of
$R_{\rm O III}$ of the various types comes from the same underlying 
population. The resultant KS probabilities are given in Table 3.
The KS test leads to the following results.
(1) The NLS1s, the BLS1s, and the S1.5s are statistically indistinguishable
in the frequency distribution of $R_{\rm O III}$.
(2) The S2$^+$ and the S2$^-$ are also statistically indistinguishable. 
(3) However, the NLS1s, of the BLS1s, and of the S1.5s have statistically 
higher $R_{\rm O III}$ values than the S2$^+$s and the S2$^-$s.
These results are consistent with the previous works;
Osterbrock et al. (1976), 
Heckman \& Balick (1979), and Shuder \& Osterbrock (1981) have mentioned that
the S1s show higher $R_{\rm O III}$ values than the S2s, and Cohen (1983)
has found that the S1.5s also show higher $R_{\rm O III}$ values than the S2s.
Nagao, Murayama, \& Taniguchi (2000a) recently confirmed that 
the observed $R_{\rm O III}$ values are statistically indistinguishable
between the NLS1s and the BLS1s (see also Rodr\'{\i}guez-Ardia et al. 2000).

It seems possible that
the difference in $R_{\rm O III}$ among the types of Seyfert galaxies is
attributed to the systematic difference in the amounts of the reddening,
because we do not make any reddening correction for the compiled emission-line
flux data. Thus we investigate the reddening effect on $R_{\rm O III}$
adopting the Cardelli's extinction curve (Cardelli, Clayton, \& Mathis 1989).
It results in that the
correction factors for the observed value of $R_{\rm O III}$ is 1.222
for the reddening if we assume $A_V$ = 1.0 mag, which is typical difference
in the amounts of reddening for NLRs
between S1s and S2s (Dahari \& De Robertis 1988;
see also De Zotti \& Gaskell 1985). This factor is too small to explain
the systematic difference of $R_{\rm O III}$ among the types of Seyfert 
galaxies. 
Therefore we conclude that the reason of the systematic difference of 
$R_{\rm O III}$ among the Seyfert types is not the extinction effect 
but any other mechanism, discussed later.

 \subsection{Correlation between $R_{\rm O III}$ and MIR-Color}

\begin{figure*}
\epsscale{1.7}
\plotone{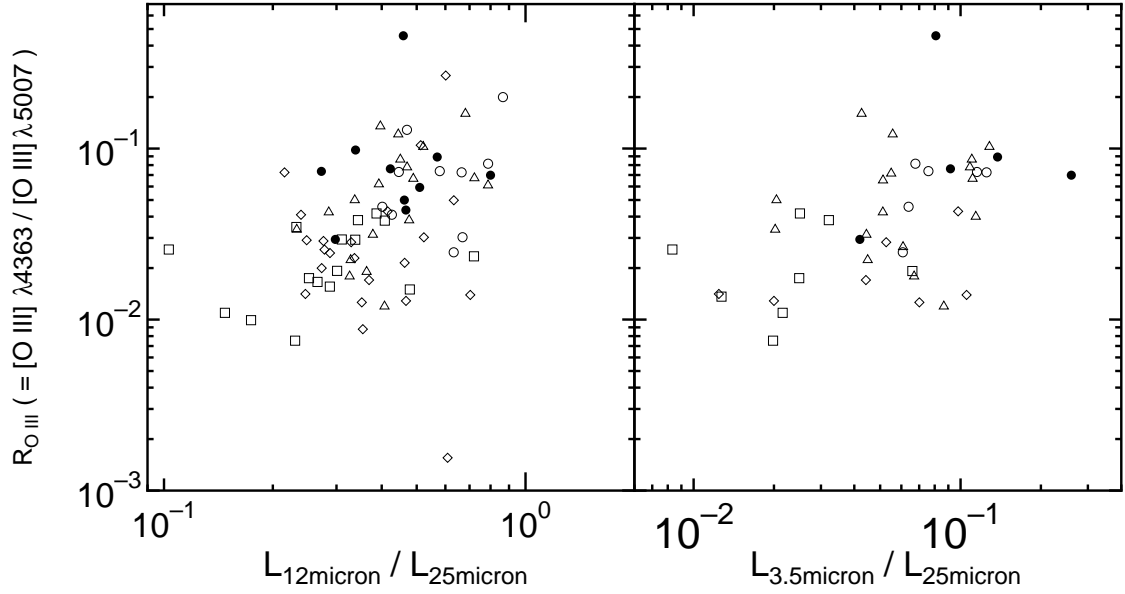}
\caption{
$R_{\rm O III}$ are plotted as functions of the flux ratios of
{\it IRAS} 12 $\mu$m (left) and of $L$-band (right) to {\it IRAS} 25 $\mu$m.
The symbols are the same as in Figure 2.
\label{fig6}}
\end{figure*}

Dust grains within dusty tori in AGNs absorb NIR to soft X-ray
photons emitted from the central engine, and re-emit the thermal radiation 
in the mid-infrared (MIR) regime.
Since the tori are quite optically thick, the MIR spectrum 
has strong dependence on the viewing angle 
(e.g., Heckman, Chanbers, \& Postman 1992; 
Giuricin, Mardirossian, \& Mezzetti 1995;
Heckman 1995; Fadda et al. 1998; Murayama, Mouri \& Taniguchi 2000). 
This means that the hot inner surface of dusty tori is seen 
when the torus is observed from a favored (i.e., more face-on) view 
but obscured when observed from a unfavored (i.e., more edge-on) view. 
Therefore, it is interesting to investigate
correlations between the MIR colors and $R_{\rm O III}$.

The dependences of $R_{\rm O III}$ on the flux ratios of
{\it IRAS} 12 $\mu$m and $L$ band to {\it IRAS} 25 $\mu$m are shown
in Figure 6. 
These two flux ratios are used to investigate the visibility of 
the hot inner surface of dusty tori in AGNs.
The method using the flux ratio of $L$ band to 
{\it IRAS} 25 $\mu$m is proposed by Murayama et al. (2000)
for the purpose of reducing the influence of star-formation.
In Figure 6, there appears a positive correlation in each diagram.
In order to investigate whether or not these positive correlations are 
statistically significant, 
we apply the Spearman's rank test (see Press et al. 1988)
where the null hypothesis is that the observed value of $R_{\rm O III}$ is not 
correlated with the flux ratios of
{\it IRAS} 12 $\mu$m and of $L$ band to {\it IRAS} 25 $\mu$m.
The resulting probabilities are 3.598 $\times 10^{-5}$ for the flux ratio
of {\it IRAS} 12 $\mu$m to {\it IRAS} 25 $\mu$m and 3.254 $\times 10^{-4}$
for the flux ratio of $L$ band to {\it IRAS} 25 $\mu$m, which mean that
the positive correlations shown in Figure 6 are statistically real.
This means that the hotter the observed MIR colors are, the higher 
the observed values of $R_{\rm O III}$ are.

 \subsection{Correlation between $R_{\rm O III}$ and HINER Components}

\begin{figure*}
\epsscale{1.7}
\plotone{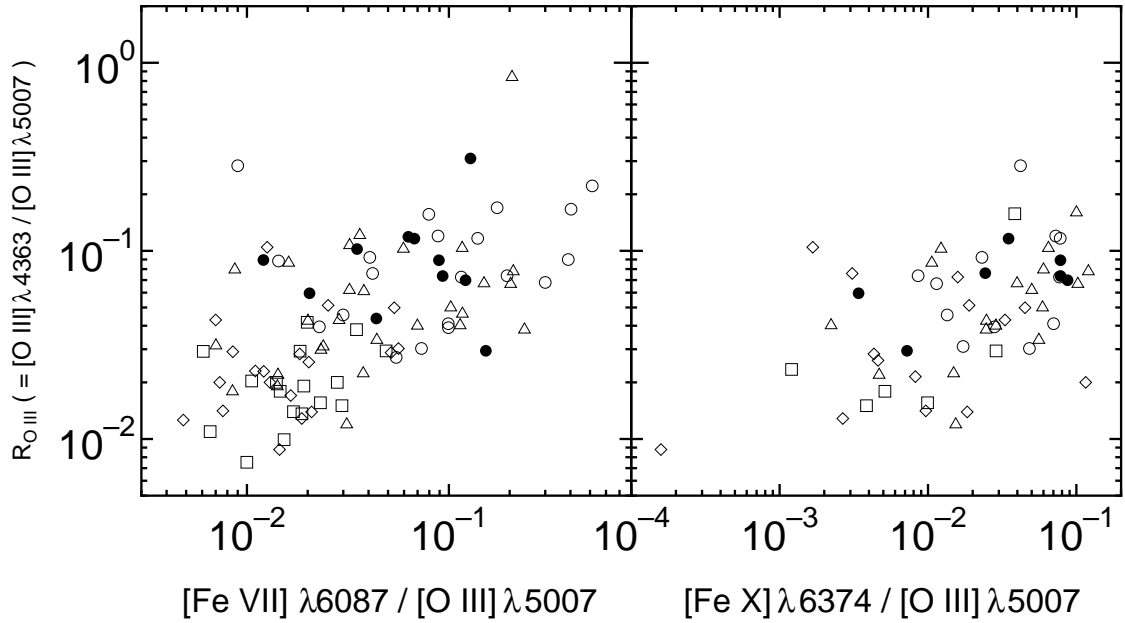}
\caption{
$R_{\rm O III}$ are plotted as functions of the line ratios of
[Fe {\sc vii}]$\lambda$6087/[O {\sc iii}]$\lambda$5007 (left) and
of [Fe {\sc x}]$\lambda$6374/[O {\sc iii}]$\lambda$5007 (right).
The symbols are the same as in Figure 2.
\label{fig7}}
\end{figure*}

Pier \& Voit (1995) investigated the hydrodynamic and line-emitting 
properties of dense clouds exposed to an AGN continuum emission at
the inner edge of the torus.
Since such regions have a large covering factor and
a high density ($n_{\rm H} \sim 10^{7-8}$ cm$^{-3}$), those clouds
are thought to be a plausible place to produce the highly-ionized
emission lines such as [Fe {\sc vii}]$\lambda$6087 and 
[Fe {\sc x}]$\lambda$6374\footnote{
The critical densities of these emission lines are 
3.6 $\times 10^7$ cm$^{-3}$ for [Fe {\sc vii}]$\lambda$6087 and 
4.8 $\times 10^9$ cm$^{-3}$ for [Fe {\sc x}]$\lambda$6374
(De Robertis \& Osterbrock 1986b).
Thus these highly ionized emission lines can be radiated in the
clouds of $n_{\rm H} \sim 10^{7-8}$ cm$^{-3}$.
However, low-ionization emission lines such as
[O {\sc iii}]$\lambda$5007 are suppressed by a collisional de-excitation in 
such a dense gas cloud because $n_{\rm cr}$ of low-ionization emission lines 
are generally low comparing to $n_{\rm H}$ of the dense gas clouds.
}.
This picture is consistent with the fact that such highly-ionized
emission lines are stronger in S1s than in S2s (Murayama \& Taniguchi 1998a;
Nagao et al. 2000c). That is, a large part of the high-ionization nuclear
emission-line regions (HINERs; Binette 1985; 
Murayama, Taniguchi, \& Iwasawa 1998) is located at the region which is
obscured in S2s by any materials, such as dusty tori.

Since $R_{\rm O III}$ is higher in S1s than in S2s
and $n_{\rm cr}$ of the [O {\sc iii}]$\lambda$4363 emission is
comparable with $n_{\rm cr}$ of [Fe {\sc vii}]$\lambda$6087, 
it is interesting to 
investigate the relation between the intensity of the 
[O {\sc iii}]$\lambda$4363 emission 
and those of the HINER emission lines, [Fe {\sc vii}]$\lambda$6087 and
[Fe {\sc x}]$\lambda$6374.
The flux of these HINER lines is normalized by the flux of 
[O {\sc iii}]$\lambda$5007 following the manner of 
Murayama \& Taniguchi (1998a) and Nagao et al. (2000c).
The results are shown in Figure 7.
There is a positive correlation for each case,
especially between $R_{\rm O III}$ and the flux ratio of
[Fe {\sc vii}]$\lambda$6087/[O {\sc iii}]$\lambda$5007. 
This means that the strong [O {\sc iii}]$\lambda$4363 emitting regions
are located at the same place as the HINERs.

It should be noted that the value of $R_{\rm O III}$ correlates to the 
intensity of [Fe {\sc x}]$\lambda$6374 worse than to that of
[Fe {\sc vii}]$\lambda$6087. This is consistent with the remark of 
Nagao et al. (2000c) that the intensity of [Fe {\sc x}]$\lambda$6374 is
less suitable to investigate the viewing angle toward tori than that of
[Fe {\sc vii}]$\lambda$6087; they claimed that a part of the
[Fe {\sc x}]$\lambda$6374 emission is radiated from spatially extended,
low-density gas (see also Korista \& Ferland 1989; Golev et al. 1995;
Murayama et al. 1998; Nagao et al. 2000b).

 \subsection{Kinematical Investigation of [O {\sc iii}] Emitting Regions}

\begin{figure*}
\epsscale{1.5}
\plotone{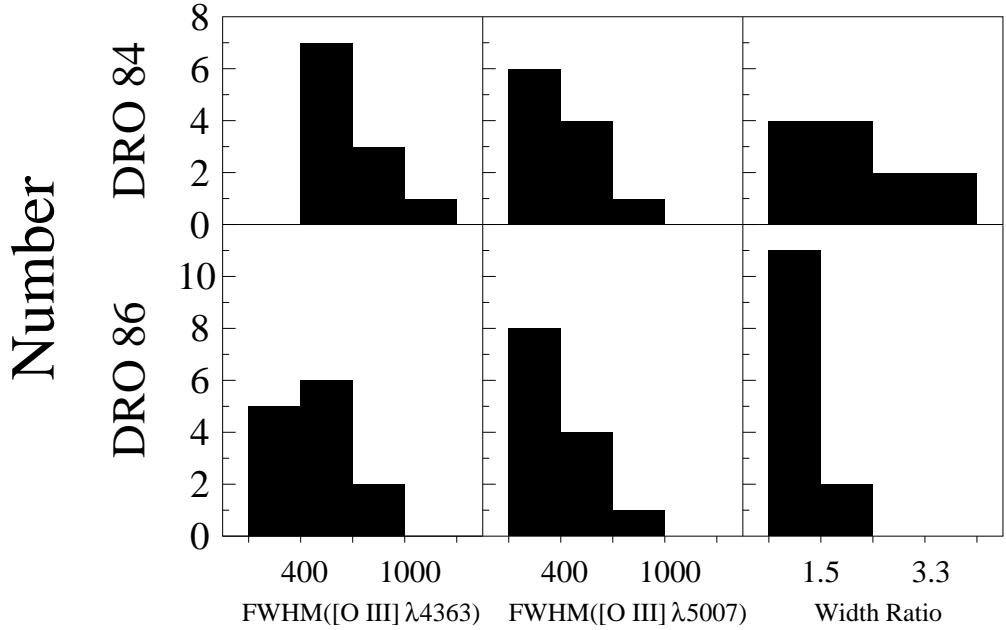}
\caption{
The frequency distributions of FWHM([O {\sc iii}]$\lambda$4363) (left),
FWHM([O {\sc iii}]$\lambda$5007) (middle), and the ratio of
FWHM([O {\sc iii}]$\lambda$4363) to FWHM([O {\sc iii}]$\lambda$5007) (right)
for the samples of DRO84, which represent broad-line Seyfert galaxies (upper),
and of DRO86, which represent narrow-line Seyfert galaxies (lower).
\label{fig8}}
\end{figure*}

\begin{figure*}
\epsscale{1.7}
\plotone{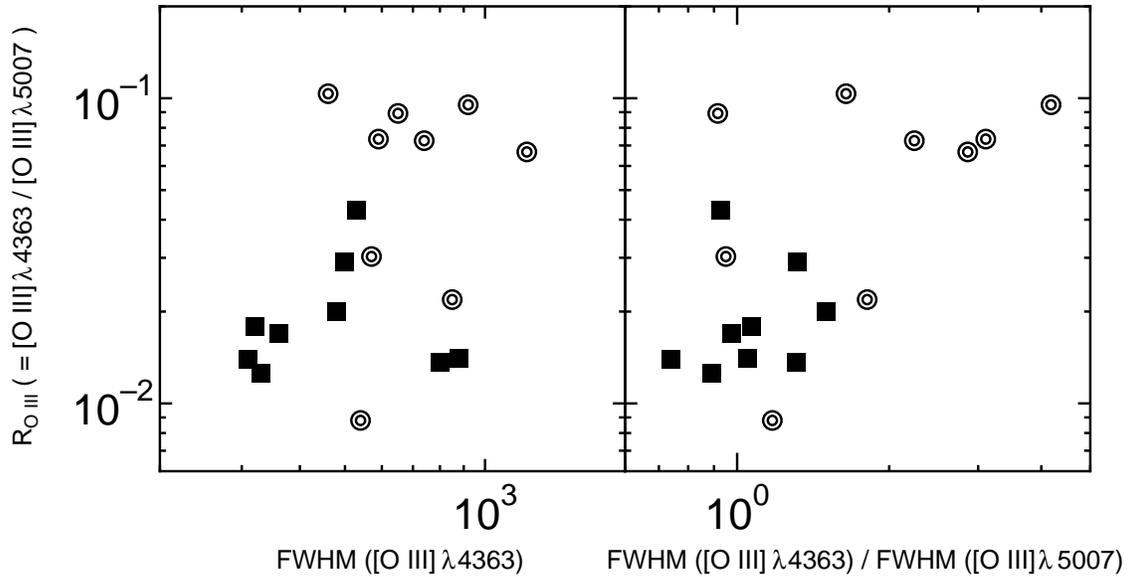}
\caption{
$R_{\rm O III}$ are plotted as functions of FWHM([O {\sc iii}]$\lambda$4363) 
(left) and the ratio of
FWHM([O {\sc iii}]$\lambda$4363) to FWHM([O {\sc iii}]$\lambda$5007) (right).
The objects in DRO84 sample and in DRO86 are shown by circles and
filled squares, respectively.
\label{fig9}}
\end{figure*}

\begin{deluxetable}{llc}
\tablenum{4}
\tablecaption{The Statistical Properties for the Sample of
   DRO84\tablenotemark{a}  and DRO86\tablenotemark{b} \label{tbl-4}}
\tablewidth{0pt}
\tablehead{
\colhead{sample} &
\colhead{} &
\colhead{mean and deviation} 
}
\startdata
DRO84& FWHM([O III]$\lambda$4363)                      & 684 $\pm$ 225 \nl
     & FWHM([O III]$\lambda$5007)                      & 387 $\pm$ 157 \nl
 &FWHM([O III]$\lambda$4363)/FWHM([O III]$\lambda$5007)& 2.03 $\pm$ 0.96 \nl
DRO86& FWHM([O III]$\lambda$4363)                      & 492 $\pm$ 168 \nl
     & FWHM([O III]$\lambda$5007)                      & 434 $\pm$ 147 \nl
 &FWHM([O III]$\lambda$4363)/FWHM([O III]$\lambda$5007)& 1.16 $\pm$ 0.25 \\
\enddata
\tablenotetext{a}{De Robertis \& Osterbrock (1984).}
\tablenotetext{b}{De Robertis \& Osterbrock (1986).}
\end{deluxetable}

\begin{deluxetable}{lc}
\tablenum{5}
\tablecaption{The Results of the KS Test Concerning
              the Kinematic Properties \label{tbl-5}}
\tablewidth{0pt}
\tablehead{
\colhead{} &
\colhead{KS Prob.} 
}
\startdata
\cutinhead{FWHM([O {\sc iii}]$\lambda$4363) versus FWHM([O {\sc iii}]$\lambda$5007)}
DRO84 & 0.012 \nl
DRO86 & 0.226 \nl
\cutinhead{DRO84 versus DRO86}
FWHM([O {\sc iii}]$\lambda$4363) & 0.023 \nl
FWHM([O {\sc iii}]$\lambda$5007) & 0.330 \nl
FWHM([O {\sc iii}]$\lambda$4363)/FWHM([O {\sc iii}]$\lambda$5007) & 0.023 \\
\enddata
\end{deluxetable}

Emission-line width of the NLR emission gives us some pieces of 
useful information
about the kinematical and geometrical properties of gas clouds in the NLRs.
Some earlier works have shown that there is a correlation between the 
emission-line width and $n_{\rm cr}$ of 
the emission line [e.g., Pelat, Fosbury, \& Alloin 1981; 
Atwood, Baldwin, \& Carswell 1982;
Filippenko \& Halpern 1984; De Robertis \& Osterbrock 1984 (DRO84); 
Filippenko 1985; De Robertis \& Osterbrock 1986 (DRO86);
Appenzeller \& \"{O}streicher 1988; Espey et al. 1994].
This correlation is broadly interpreted as follows.
A given emission line is emitted most efficiently from gas clouds whose 
densities are close to $n_{\rm cr}$. 
On the other hand, we can use line width as a rough measure of location of 
the emitting region if we assume that the NLR line widths are dominated 
either by random virialized motion or by Keplerian rotation.
Therefore the correlation between line width and $n_{\rm cr}$ suggests that
high-density gas clouds are located near the central engine relative to 
low-density gas clouds (DRO84; DRO86; see also Ferguson et al. 1997). 
This allows us to study the geometrical relationship between the 
[O {\sc iii}]$\lambda$4363 and the [O {\sc iii}]$\lambda$5007 
emitting regions. 

DRO84 measured the line widths of [O {\sc iii}]$\lambda$4363 and 
[O {\sc iii}]$\lambda$5007 for 11 broad-line Seyfert galaxies 
(NLS1s, BLS1s, and S1.5s) and DRO86 measured those for
13 narrow-line Seyfert galaxies (S2$^+$s and S2$^-$s).
Using these data, which are corrected for the instrumental broadening,
we compare the kinematical and geometrical properties 
between the [O {\sc iii}]$\lambda$4363 and 
the [O {\sc iii}]$\lambda$5007 emitting regions in the two samples.
Note that the [O {\sc iii}]$\lambda$4363 emission is weak and that
sometimes the deblending this line from H$\gamma$ may be difficult.
Therefore we do not attempt to collect the line-width data from 
a large number of the literature and use only ones presented by 
De Robertis \& Osterbrock.
However, the difficulty in the measurement of the line widths 
may cause any systematical errors, which must be kept in mind.
Since the numbers of the samples of DRO84 and DRO86 are small, 
we do not divide the sample into more detailed ones 
in this section.

The means and the 1$\sigma$ deviations of the full-width at half maximum (FWHM)
of [O {\sc iii}]$\lambda$4363 and [O {\sc iii}]$\lambda$5007, 
and ratios of them are given in Table 4. 
The histograms of these parameters are shown in Figure 8.
In order to investigate whether or not the distributions of the line width of 
[O {\sc iii}]$\lambda$4363 and that of [O {\sc iii}]$\lambda$5007 are
statistically different, and whether or not these distributions are
statistically different between the samples of DRO84 and DRO86,
we apply the KS test. 
The KS test leads to the following results.
(1) For the DRO84 sample, FWHM([O {\sc iii}]$\lambda$4363) is larger than
FWHM([O {\sc iii}]$\lambda$5007) though the statistical significance
is low ($P_{\rm KS}$ = 0.012).
(2) For the DRO86 sample, FWHM([O {\sc iii}]$\lambda$4363) and
FWHM([O {\sc iii}]$\lambda$5007) are statistically indistinguishable
($P_{\rm KS}$ = 0.226).
(3) FWHM([O {\sc iii}]$\lambda$4363) of the DRO84 sample is larger than
that of the DRO86 sample though the statistical significance
is low ($P_{\rm KS}$ = 0.023).
(4) FWHM([O {\sc iii}]$\lambda$5007) of the DRO84 sample and that of the
DRO86 sample are statistically indistinguishable ($P_{\rm KS}$ = 0.330).
And finally, (5) the ratio of 
FWHM([O {\sc iii}]$\lambda$4363)/FWHM([O {\sc iii}]$\lambda$5007) of the
DRO84 sample is larger than that of the DRO86 sample though the statistical 
significance is low ($P_{\rm KS}$ = 0.023).
These results are summarized in Table 5.
All these results support the idea that the strong [O {\sc iii}]$\lambda$4363
emitting region is located at inner region comparing to 
the [O {\sc iii}]$\lambda$5007
emitting region, and such a strong [O {\sc iii}]$\lambda$4363 
emitting region is visible only in S1s but obscured in S2s,
although a much larger sample will be necessary to confirm these arguments.

To investigate the relationship between the visibility and kinematics
of the [O {\sc iii}]$\lambda$4363 emitting regions more directly, 
we examine the dependence
of $R_{\rm O III}$ on FWHM([O {\sc iii}]$\lambda$4363) and that on the ratio of
FWHM([O {\sc iii}]$\lambda$4363)/FWHM([O {\sc iii}]$\lambda$5007) in Figure 9.
In order to examine whether or not there are any correlations in these
parameters statistically, we apply the Spearman's rank test
where the null hypothesis is that the observed value of $R_{\rm O III}$ is not 
correlated with FWHM([O {\sc iii}]$\lambda$4363) or
FWHM([O {\sc iii}]$\lambda$4363)/FWHM([O {\sc iii}]$\lambda$5007).
The resulting probabilities are 0.165 for FWHM([O {\sc iii}]$\lambda$4363) and
0.032 for FWHM([O {\sc iii}]$\lambda$4363)/FWHM([O {\sc iii}]$\lambda$5007).
These results mean that there is no correlation between $R_{\rm O III}$ and
FWHM([O {\sc iii}]$\lambda$4363) while there is a marginal tendency of
a positive correlation between $R_{\rm O III}$ and the ratio of
FWHM([O {\sc iii}]$\lambda$4363)/FWHM([O {\sc iii}]$\lambda$5007).
This difference is caused because FWHM([O {\sc iii}]$\lambda$4363)
reflects not only the location of the [O {\sc iii}]$\lambda$4363 emitting
region but also the mass of a supermassive black hole
while the effect of the dispersion of the mass of a supermassive black hole
among objects is reduced in the value of
FWHM([O {\sc iii}]$\lambda$4363)/FWHM([O {\sc iii}]$\lambda$5007).

It should be noted that there are some lines of evidence which show that 
some of broader emission-line widths of highly-ionized emission lines are
due to outflows, not to the depth of gravitational potentials
(e.g., Moore \& Cohen 1996; Kaiser et al. 2000; Nelson et al. 2000;
Crenshaw \& Kraemer 2000). 
If this is the case, line widths might not contain the information concerning
the geometry of line-emitting gas clouds. However, unfortunately, 
the present data cannot distinguish these two interpretations.

\section{DISCUSSION}

 \subsection{Where is the [O {\sc iii}]$\lambda$4363 Emitting Region in AGNs?}

\begin{figure*}
\epsscale{0.9}
\plotone{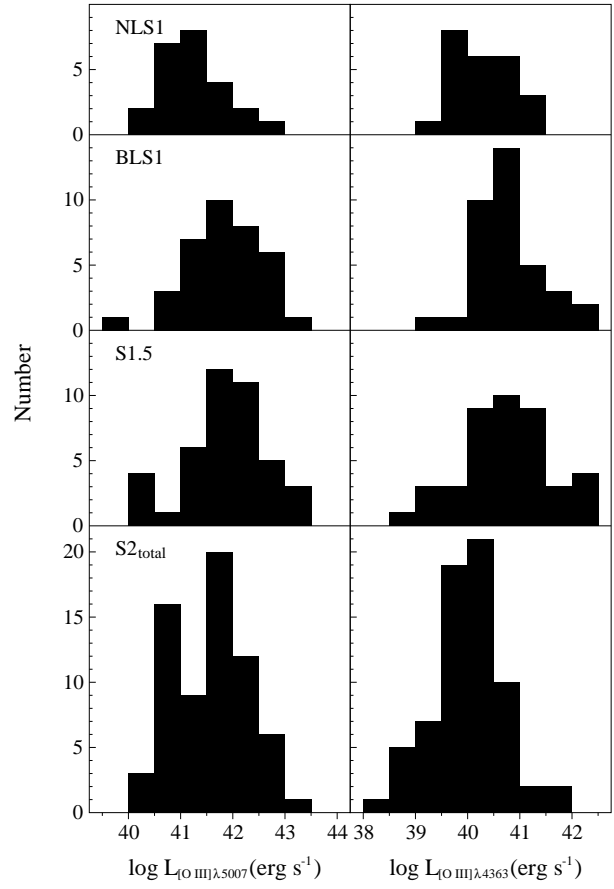}
\caption{
The frequency distributions of the [O {\sc iii}]$\lambda$5007 luminosity 
(left) and the [O {\sc iii}]$\lambda$4363 luminosity (right) 
for the NLS1s, the BLS1s, the S1.5s, and the S2$_{\rm total}$s.
\label{fig10}}
\end{figure*}

\begin{deluxetable}{lcccc}
\tablenum{6}
\tablecaption{The Results of the KS Test Concerning the
              $L_{\rm [O III]\lambda 5007}$ and 
              $L_{\rm [O III]\lambda 4363}$ \label{tbl-6}}
\tablewidth{0pt}
\tablehead{
\colhead{Type} &
\colhead{NLS1} &
\colhead{BLS1} & 
\colhead{S1.5} & 
\colhead{S2$_{\rm total}$}
}
\startdata
\cutinhead{$L_{\rm [O III]\lambda 5007}$}
NLS1             & \nodata & 4.291 $\times 10^{-3}$ & 4.394 $\times 10^{-4}$ & 4.445 $\times 10^{-2}$ \nl
BLS1             & \nodata & \nodata                & 9.312 $\times 10^{-1}$ & 8.359 $\times 10^{-2}$ \nl
S1.5             & \nodata & \nodata                & \nodata                & 1.225 $\times 10^{-2}$ \nl
S2$_{\rm total}$ & \nodata & \nodata                & \nodata                & \nodata                \nl
\cutinhead{$L_{\rm [O III]\lambda 4363}$}
NLS1             & \nodata & 2.519 $\times 10^{-2}$ & 2.263 $\times 10^{-2}$ & 4.419 $\times 10^{-1}$ \nl
BLS1             & \nodata & \nodata                & 7.925 $\times 10^{-1}$ & 2.515 $\times 10^{-5}$ \nl
S1.5             & \nodata & \nodata                & \nodata                & 5.194 $\times 10^{-5}$ \nl
S2$_{\rm total}$ & \nodata & \nodata                & \nodata                & \nodata                \\
\enddata
\end{deluxetable}

In this section, we discuss why the observed values of $R_{\rm O III}$ are
higher in S1s than in S2s and
how the high $R_{\rm O III}$ comparing to the predicted values by
simple photoionization models is achieved.

The first problem is the type dependence of the observed values of 
$R_{\rm O III}$, shown in Figure 5. 
There are two possible interpretations to understand 
this dependence. One is that the [O {\sc iii}]$\lambda$5007 emission is
stronger in S2s than S1s due to the intrinsically (i.e., not due to 
inclination effects) larger size of NLRs of S2s as proposed by 
Schmitt \& Kinney (1996). 
Another is that the strong 
[O {\sc iii}]$\lambda$4363 emitting regions exist somewhere but obscured
by something on the line of sight when we see S2s.
Because the intrinsic AGN luminosity is similar among the various types of
Seyferts in our sample as mentioned in Section 2.2, the former case predicts
the stronger [O {\sc iii}]$\lambda$5007 luminosity in S2s than in S1s and 
similar [O {\sc iii}]$\lambda$4363 luminosity among the Seyfert types.
On the other hand, the latter case predicts the similar 
[O {\sc iii}]$\lambda$5007 luminosity among the Seyfert types and the stronger
[O {\sc iii}]$\lambda$4363 luminosity in S1s than in S2s.
These two cases may be the extreme ones and the real situation 
might be intermediate between the two cases.
However, it is interesting to investigate which case is close to 
the observed properties of emission-line spectra. 
Therefore we compare the observed emission-line luminosity of 
[O {\sc iii}]$\lambda$5007 and [O {\sc iii}]$\lambda$4363 among various 
types of Seyferts, which is shown in Figure 10.
To quantify the statistical significance of the difference in 
the luminosity distribution among the Seyfert types, we apply the KS test 
where the null hypothesis is that the distribution of these emission-line
luminosities come from the same underlying population.
The KS test leads to the following results. 
(1) The distributions of the [O {\sc iii}]$\lambda$5007 luminosity are 
statistically indistinguishable among the BLS1s, the S1.5s, and the S2s 
though the [O {\sc iii}]$\lambda$5007 luminosity of the NLS1s are 
weaker than that of other Seyfert types. 
(2) The distributions of the [O {\sc iii}]$\lambda$4363 luminosity are 
statistically indistinguishable among the NLS1s, BLS1s, and the S1.5s 
though the [O {\sc iii}]$\lambda$4363 luminosity of the S2s are 
weaker than that of the BLS1s and the S1.5s.
The KS probabilities are given in Table 6.
These results are consistent with the latter case; i.e., 
the type dependence of the observed values of $R_{\rm O III}$ is
not due to the dilution by a more extended NLR in S2s but due to the
enhancement of the [O {\sc iii}]$\lambda$4363 emission in S1s.
The reason why the [O {\sc iii}]$\lambda$5007 luminosity is smaller 
in the NLS1s than in other types may be that NLS1s are the extreme objects
on the ``eigenvector 1'' of Boroson \& Green (1992).
More explicitly, there is a relation that 
the weaker the [O {\sc iii}]$\lambda$5007 emission is, the narrower the
FWHM of H$\beta$ is (e.g., Boroson \& Green 1992; 
Brandt \& Boller 1998; Sulentic et al. 2000).
Following this relation, NLS1s may tend to exhibit 
weak [O {\sc iii}]$\lambda$5007 emissions as shown in Figure 10.

The properties of emission-line width described in Section 3.4 also support 
the idea that the dependence of $R_{\rm O III}$ on AGN types is not due to
the dilution of $R_{\rm O III}$ by extended low density gas in S2s but due to
the obscuration of the strong [O {\sc iii}]$\lambda$4363 emitting region
in S2s. The S1s exhibit the broad 
[O {\sc iii}]$\lambda$4363 comparing to [O {\sc iii}]$\lambda$5007
while the S2s do not so. This suggests the existence of the inner, strong
[O {\sc iii}]$\lambda$4363 emitting region which is obscured in S2s.
We, therefore, conclude that the dependence of $R_{\rm O III}$ on AGN types
is attributed to the obscuration effect.

Now we consider the following problems. 
{\it Where is the [O {\sc iii}]$\lambda$4363 emitting region in AGNs?} 
And, {\it how is the high $R_{\rm O III}$ comparing 
to the predicted values by simple photoionization models achieved?}
According to the current unified model of AGNs, it is natural to consider 
that the material obscuring the strong [O {\sc iii}]$\lambda$4363 
emitting regions in S2s is dusty tori. If this is the case,
the strong [O {\sc iii}]$\lambda$4363 emitting region may be either
the inner surface of dusty tori described by Pier \& Voit (1995) or
the dense gas clouds near the central engine, which are obscured by the tori,
though these two alternatives cannot be distinguished only by the 
statistical tests presented in this paper.
This is consistent with the similarity of the location between the HINER and
the strong [O {\sc iii}]$\lambda$4363 emitting region, which is
suggested by the correlation between the intensity of the HINER emission and 
$R_{\rm O III}$ (see Section 3.3)
because large parts of the HINER emission is thought to arise from 
such dense clouds (Pier \& Voit 1995; Murayama \& Taniguchi 1998a, 1998b; 
Nagao et al. 2000c).
The MIR properties described in Section 3.2 also support this geometrical 
consideration of the [O {\sc iii}]$\lambda$4363 emitting region; i.e.,
the correlations between the MIR colors and $R_{\rm O III}$ mean the
similarity between the visibility of relatively hot inner surface of 
dusty tori and that of the strong [O {\sc iii}]$\lambda$4363 emitting region.

However, some of S2s also exhibit large $R_{\rm O III}$, which is also 
difficult to be explained by the simple one-zone photoionization model.
They may be have dense gas clouds in NLR, which is not obscured 
by dusty tori because of the large distance from the nucleus, 
or the escaping [O {\sc iii}]$\lambda$4363 emission 
from the leaky parts of the tori.

 \subsection{Interpretation Using the Dual-Component Photoionization Model}

\begin{figure*}
\epsscale{1.8}
\plotone{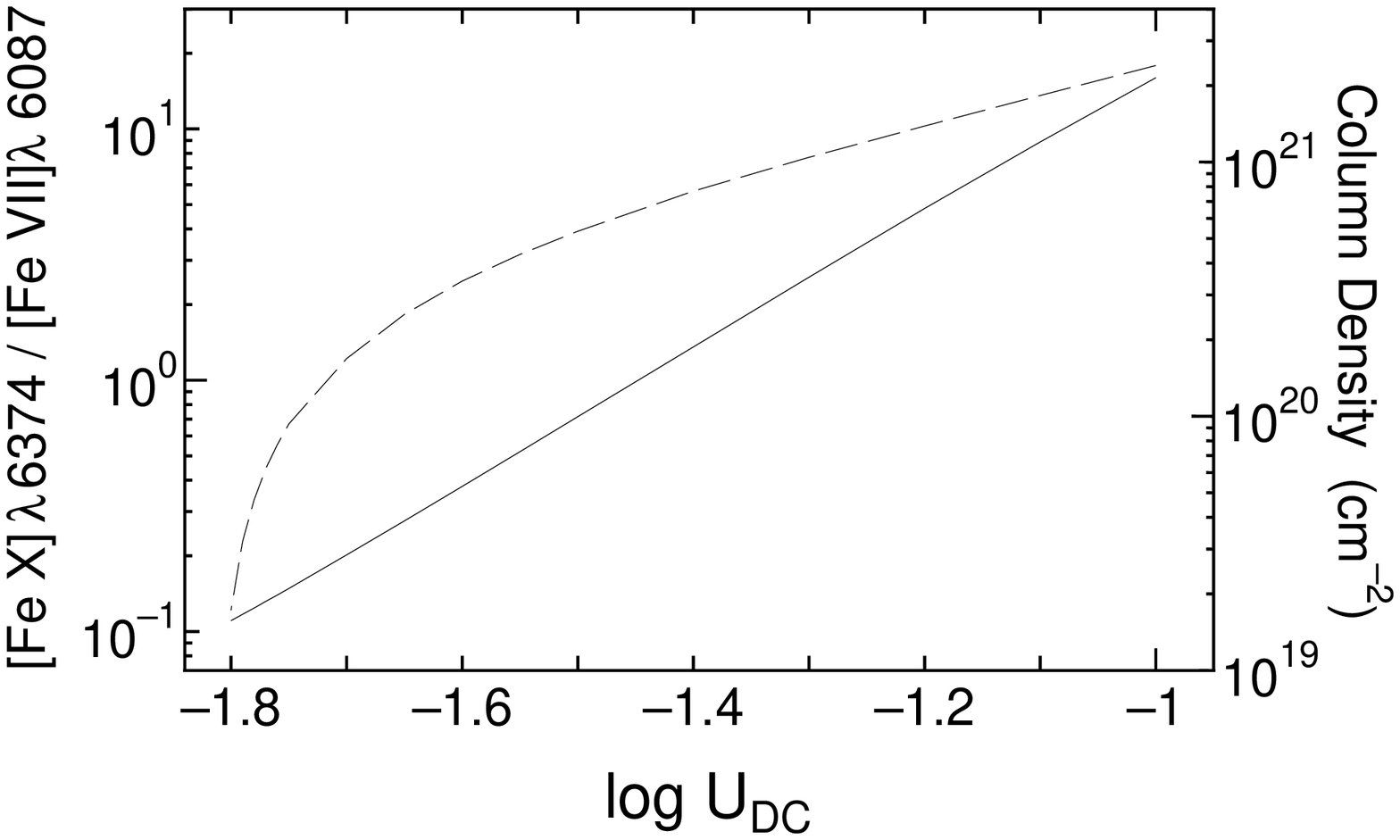}
\caption{
The calculated emission-line ratio of 
[Fe {\sc x}]$\lambda$6374/[Fe {\sc vii}]$\lambda$6087 (solid line)
and the column density (dashed line)
of the truncated torus component are shown as functions of the ionization 
parameter of the torus component, $U_{\rm DC}$.
\label{fig11}}
\end{figure*}

\begin{figure*}
\epsscale{1.0}
\plotone{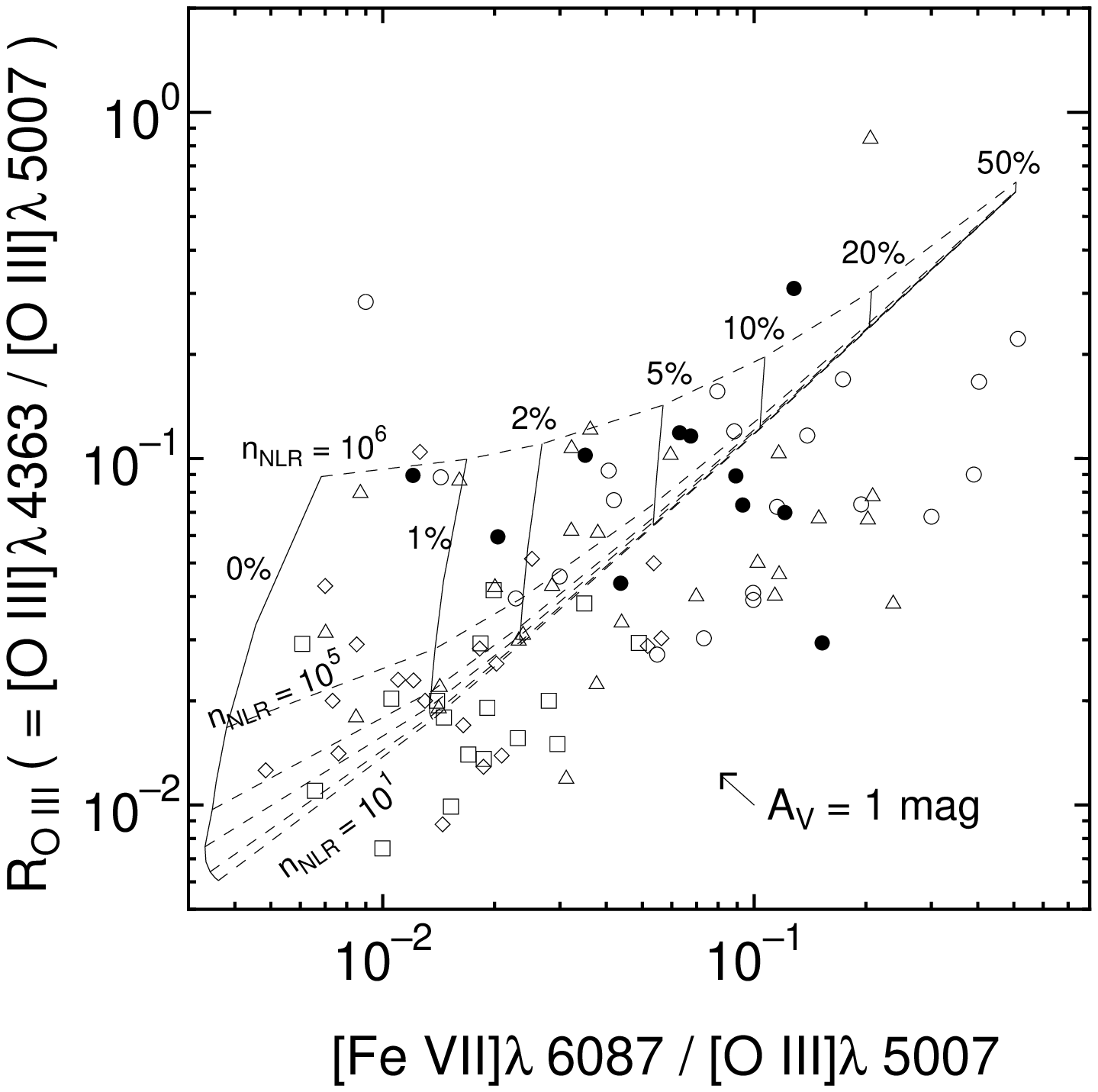}
\caption{
The diagram of $R_{\rm O III}$ versus
[Fe {\sc vii}]$\lambda$4363/[O {\sc iii}]$\lambda$5007. 
The symbols are the same as in Figure 2.
Our model calculations are superposed in the figure.
The fraction of the contribution from torus components are also shown.
The data points will move on the diagram as shown by the arrow 
if the extinction correlation of $A_V$ = 1.0 is applied.
\label{fig12}}
\end{figure*}

As described in Section 4.1, high-density gas clouds obscured by 
tori, which are located either at the inner surface of dusty tori or
near the central engine where is obscured by the tori,
are thought to emit a large fraction of 
the [O {\sc iii}]$\lambda$4363 emission.
In order to investigate whether or not such an idea is consistent with
photoionization scenarios quantitatively, 
we perform dual-component photoionization model
calculations following the manner of Murayama \& Taniguchi (1998b).
This method takes account of such high-density gas clouds
as a strong [O {\sc iii}]$\lambda$4363 emitter, 
in addition to the typical NLR component.
Here we assume the second situation, i.e., 
the strong [O {\sc iii}]$\lambda$4363 emitting region is not physically
associated by the tori. Therefore we do not consider effects of
dust, such as depletion of heavy metals, 
through the following calculations.

Our calculation methods are as follows.
We perform photoionization model calculations using the 
spectral synthesis code {\it Cloudy} version 90.04 (Ferland 1996), 
which solves the 
equations of statistical and thermal equilibrium and produces a 
self-consistent model of the run of temperature as a function of depth into 
the nebula. Here we assume an uniform-density gas cloud with 
a plane-parallel geometry.
The dense component (DC) is assumed to be truncated clouds; i.e.,
optically thin clouds for the ionizing photons, for the purpose of
avoiding unusually strong [O {\sc i}] emission 
(see Murayama \& Taniguchi 1998b).
The parameters for the calculations are (1) the hydrogen density of 
the cloud ($n_{\rm DC}$ and $n_{\rm NLR}$), 
(2) the ionization parameter ($U_{\rm DC}$ and $U_{\rm NLR}$), 
which is defined 
as the ratio of the ionizing photon density to the electron density,
(3) the thickness of the torus component which is represented
by the optical depth for ionizing photons,
(4) the chemical compositions of the gas, 
(5) the shape of the input SED of ionizing photons, and
(6) the fraction of DC to the NLR component.

Here we assume $n_{\rm DC} = 10^7$ cm$^{-3}$. 
We perform several model runs covering 
10$^1$ cm$^{-3} \leq n_{\rm NLR} \leq 10^6$ cm$^{-3}$.
The ionization parameter of the NLR component is assumed as 
$U_{\rm NLR} = 10^{-2}$. The ionization parameter and the hydrogen column 
density of DC are determined using following two conditions;
([Fe {\sc x}]$\lambda$6374/[Fe {\sc vii}]$\lambda$6087)$_{\rm DC}$ = 0.8
and ([Fe {\sc vii}]$\lambda$6087/[O {\sc iii}]$\lambda$5007)$_{\rm DC}$
= 1.0. The former ratio is the typical value of Seyfert galaxies
(Nagao et al. 2000c) and the latter condition is introduced by 
Murayama \& Taniguchi (1998b) as a truncated dense gas cloud.
As the result, $U_{\rm DC} = 10^{-1.48}$ and 
$N_{\rm DC} = 10^{20.76}$ cm$^{-2}$ are adopted 
\footnote{
Because it is known that there is uncertainty in collision strengths 
for the [Fe {\sc x}]$\lambda$6374 emission, the derived values of
$U_{\rm DC}$ and $N_{\rm DC}$ may also suffer such uncertainties.
}
(see Figure 10).
The calculations are stopped when the gas temperature falls to 4000 K 
for the NLR component.
We set the gas-phase elemental abundances to be solar ones taken from
Grevesse \& Anders (1989) with extensions by Grevesse \& Noels (1993).
We adopt the power-law continuum as the input spectrum, where
the spectral index is assumed as $\alpha$ = --1.5 (see Ferland \& Netzer 1983)
between 10 $\mu$m and 50 keV for the form $f_{\nu} \propto \nu^{\alpha}$.
The spectral index is set to $\alpha$ = 2.5 at lower energy 
(i.e., $\lambda \geq 10 \mu$m) and to $\alpha$ = --2 at higher energy
(i.e., $h\nu \geq$ 50 keV).
The fraction of DC to the NLR component is treated 
as a free parameter in our calculations.
The further details for this dual-component photoionization model are
described in Murayama (1998) and Murayama \& Taniguchi (1998b).

We present our results of model calculations
and compare them with the observations
in Figure 12, which is a diagram of $R_{\rm O III}$ versus 
[Fe {\sc vii}]$\lambda$6087/[O {\sc iii}]$\lambda$5007.
We find that the model grids are roughly consistent with the observations
if we take the effects of the correction for the extinction into account.
Though the dispersion of observation is larger than the model grids,
this is thought to be attributed the fact that the parameters, such as 
$U_{\rm DC}$, $U_{\rm NLR}$, $n_{\rm DC}$, and $N_{\rm DC}$,
are different from object to object.
It is shown that the $R_{\rm O III}$ of the S1s can be explained by 
introducing a 5\% $\sim$ 20\% contribution from DC while 
the $R_{\rm O III}$ of the S2s can be explained by introducing 
a 0\% $\sim$ 2\% contribution from DC.
These fractions are consistent with the results of 
Murayama \& Taniguchi (1998b), who introduce a $\sim 10 \%$ contribution from 
the dense gas clouds to explain intensities of the HINER emission of S1s.

 \subsection{$R_{\rm O III}$ in LINERs}

\begin{figure*}
\epsscale{0.6}
\plotone{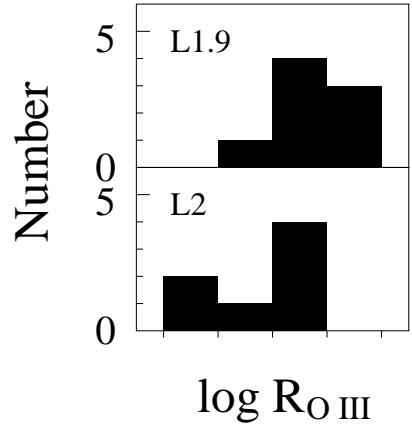}
\caption{
The frequency distributions of $R_{\rm O III}$ for the L1.9s and the L2s.
\label{fig13}}
\end{figure*}

\begin{figure*}
\epsscale{1.7}
\plotone{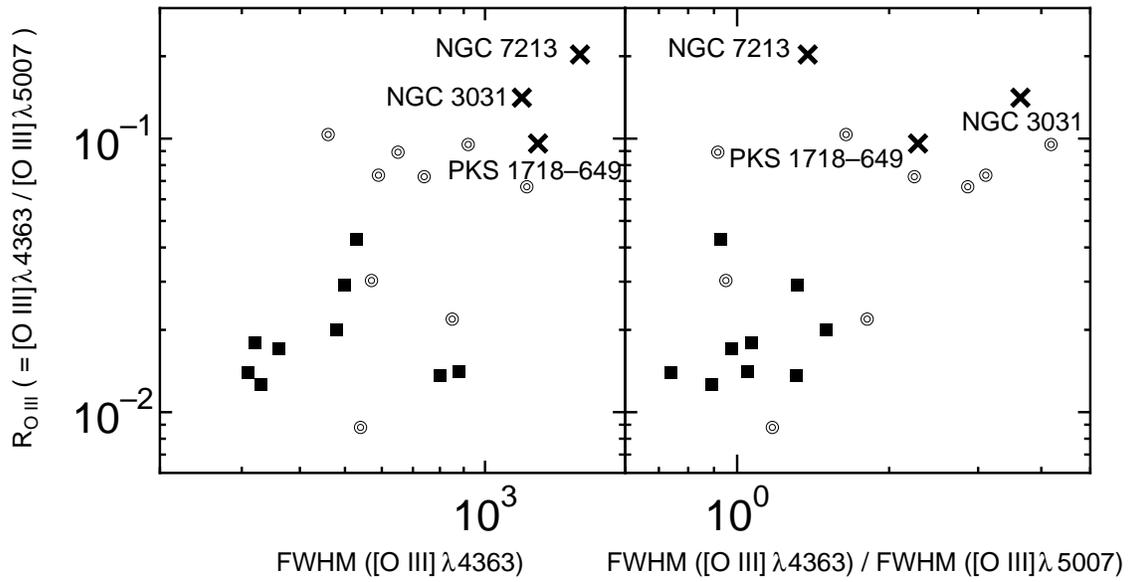}
\caption{
Same as Figure 9 but the sample of LINERs (NGC 3031, NGC 7213, and
PKS 1718-649) are added. The crosses denote the LINERs.
\label{fig14}}
\end{figure*}

\begin{deluxetable}{lll}
\tablenum{7}
\tablecaption{Observed values of $R_{\rm OIII}$ for LINERs \label{tbl-7}}
\tablehead{
\colhead{Object Name} &
\colhead{$R_{\rm OIII}$} &
\colhead{Reference\tablenotemark{a}} 
}
\startdata
\cutinhead{LINER with a broad component}
NGC 1052    & 0.0606 & HFS93 \nl
            & 0.0353 & HFS97 \nl
NGC 1275    & 0.0923 & HFS93 \nl
NGC 3031    & 0.1884 & HFS93 \nl
            & 0.0938 & HFS96 \nl
NGC 3226    & 0.0901 & HFS97 \nl
NGC 4278    & 0.1353 & HFS97 \nl
NGC 4395    & 0.0312 & HFS93 \nl
NGC 4579    & 0.0510 & HFS97 \nl
NGC 7213    & 0.2030 & FH84  \nl
\cutinhead{LINER without a broad component}
NGC 1167    & 0.0090 & HFS93 \nl
NGC 1961    & 0.0750 & HFS97 \nl
NGC 3504    & 0.0618 & HFS93 \nl
NGC 4102    & 0.0300 & HFS97 \nl
NGC 6500    & 0.0500 & HFS97 \nl
NGC 7714    & 0.0121 & HFS93 \nl
PKS 1718-649& 0.0960 & F85   \\
\enddata
\tablenotetext{a}{
References for the data of $R_{\rm OIII}$.
Each abbreviation means as follows; 
F85: Filippenko (1985);
FH84: Filippenko \& Halpern (1984);
HFS93: Ho, Filippenko, \& Sargent (1993);
HFS96: Ho, Filippenko, \& Sargent (1996); and
HFS97: Ho, Filippenko, \& Sargent (1997a).
}
\end{deluxetable}

In some low-ionization nuclear emission-line regions (LINERs),
the observed values of $R_{\rm O III}$ are far larger than that predicted by
one-zone photoionization models. Because of this property, the dominant
mechanism for the ionization in LINERs has been frequently regarded as shock
ionization (e.g., Fosbury et al. 1978; Heckman 1980; 
Baldwin, Phillips, \& Terlevich 1981). 
However, Filippenko (1985) pointed out that 
there is a correlation between the 
emission-line width and $n_{\rm cr}$ of the emission line 
over the range 10$^3$ cm$^{-3} \leq n_{\rm cr} \leq 10^7$ cm$^{-3}$
in a LINER PKS 1718-649.
This suggests that LINERs may also possess high-density regions up to 10$^7$
cm$^{-3}$, which mean that the high $R_{\rm O III}$ in LINERs may be explained
by photoionization models.

Although it is not clear whether or not there is a dusty torus in all LINERs,
there is several pieces of evidence that the unified models of AGNs can
apply to some of LINERs; i.e., some LINERs exhibit broad components
in their optical spectra (e.g., Ho et al. 1997b), 
in UV spectra (e.g., Barth et al. 1996), 
or only in polarized spectra (Barth, Filippenko, \& Moran 1999a, 1999b).
Therefore it is interesting to examine the properties of $R_{\rm O III}$ 
of LINERs in the framework of our dual component model.

In order to investigate this issue, we compiled $R_{\rm O III}$ of LINERs
from the literature. 
Since the optical spectra of LINERs are often
contaminated by stellar features from the host galaxy strongly, 
careful subtraction of
such stellar features from observed spectra is needed to discuss 
the properties of faint emission lines such as [O {\sc iii}]$\lambda$4363.
Our compiled sample consists of the objects that such careful subtraction 
was applied to;
8 LINERs with broad components (L1.9s) and 7 LINERs without
broad components (L2s).
Here it must be kept in mind that some of the L2s may not be AGNs;
a part of LINERs may be shock-heated galaxy 
(e.g., Heckman 1980; Baldwin et al. 1981; Heckman 1986; 
Gonz\'arez Delgado \& P\'erez 1996) and others may be the objects
ionized by hot stellar component (e.g., Filippenko \& Terlevich 1992; 
Binette et al. 1994; Alonso-Herrero et al. 2000; 
Taniguchi, Shioya, \& Murayama 2000).
It is noted that the compiled samples may be biased in favor of the higher
$R_{\rm O III}$ objects because it is often difficult to detect weak 
[O {\sc iii}]$\lambda$4363 emission
owing to the relatively strong stellar feature.
The objects we compiled are given in Table 7, and the histograms of 
$R_{\rm O III}$ for the L1.9s and the L2s are shown in Figure 13.
The means and the deviations of $R_{\rm O III}$ are 0.099 $\pm$ 0.054
for the L1.9s and 0.048 $\pm$ 0.030 for the L2s.
There are a tendency that the values of the L1.9s are higher than those of L2s
though these are statistically indistinguishable ($P_{\rm KS}$ = 0.252).

In the sample, the emission-line width of [O {\sc iii}]$\lambda$4363
has been measured for three LINERs; NGC 3031 
(Ho, Filippenko, \& Sargent 1996), NGC 7213 (Filippenko \& Halpern 1984), 
and PKS 1718-649 (Filippenko 1985). 
These data follow our dual component model; i.e.,
higher $R_{\rm O III}$ objects show larger values of 
FWHM([O {\sc iii}]$\lambda$4363) and 
FWHM([O {\sc iii}]$\lambda$4363)/FWHM([O {\sc iii}]$\lambda$5007 (Figure 14).

These results seem to suggest that the [O {\sc iii}]$\lambda$4363 emitting
regions are located at inner than the [O {\sc iii}]$\lambda$5007 emitting
regions and that the [O {\sc iii}]$\lambda$4363 emission has an anisotropic
property for LINERs, too. 
Further observations are needed to investigate this issue in detail.

\section{SUMMARY}

In this paper we proposed the idea that a large fraction of 
[O {\sc iii}]$\lambda$4363 originates in dense gas clouds 
obscured by the torus, which cause high values of $R_{\rm O III}$
comparing to predictions of simple one-zone photoionization models.
We have shown some observational properties of $R_{\rm O III}$ which support
our model.
  \begin{itemize}
    \item {\it The values of $R_{\rm O III}$ of the NLS1s, the BLS1s, and 
          the S1.5s are higher than those of the S2s.} 
          This difference suggests a large fraction of
          [O {\sc iii}]$\lambda$4363 emission is hidden by the torus in S2s.
    \item {\it The higher-$R_{\rm O III}$ objects show hotter MIR colors.}
          The hotter MIR colors are thought to be attributed 
          to the hotter dusty grains located at the inner surface of 
          the dusty tori, which can be seen if we see the torus from 
          more face-on view.
          Therefore, this means that the higher $R_{\rm O III}$ objects are
          seen from more face-on view toward dusty tori than the lower 
          $R_{\rm O III}$ objects.
    \item {\it The higher-$R_{\rm O III}$ objects show stronger HINER 
          emission.} 
          Since a large fraction of HINER emission is thought to arise from
          dense gas clouds at the inner surface of the dusty tori 
          (Murayama \& Taniguchi 1998a, 1998b),
          this means that the higher $R_{\rm O III}$ can be attributed to
          the significant flux contribution from such a high dense cloud
          as described by Pier \& Voit (1995).
    \item {\it The S1s have wider FWHM([O {\sc iii}]$\lambda$4363) and
          FWHM([O {\sc iii}]$\lambda$4363)/FWHM([O {\sc iii}]$\lambda$5007)
          than the S2s.} This suggests that the [O {\sc iii}]$\lambda$4363
          emitting regions are located inner than the 
          [O {\sc iii}]$\lambda$5007 emitting regions and have 
          an anisotropic property.
    \item {\it The higher-$R_{\rm O III}$ objects show larger
          FWHM([O {\sc iii}]$\lambda$4363)/FWHM([O {\sc iii}]$\lambda$5007)
          ratios.}
          This also suggests that the [O {\sc iii}]$\lambda$4363 emitting 
          regions are located inner than the [O {\sc iii}]$\lambda$5007 
          emitting regions.
  \end{itemize}

As shown in section 3.4, there is too little information to discuss the
kinematical and structural properties of the [O {\sc iii}]$\lambda$4363
emitting region because it is often difficult to observe this emission line 
accurately.
Therefore further observations are needed to confirm this dual-component
model.
In particular, the spatial distribution of $R_{\rm O III}$ should be 
investigated to judge the validity for this model.
This dual-component model predicts that the higher $R_{\rm O III}$ ($\sim$ 0.1)
is seen only in nuclear region.

\acknowledgments

We would like to thank Gary Ferland for providing his code $Cloudy$ 
to the public. We also thank the anonymous referee and Yasuhiro Shioya 
for some useful comments, and Shingo Nishiura for his kind assistance.
YT would like to thank Rolf-Peter Kudritzki, Bob McLaren, and Dave Sanders
at Institute for Astronomy, University of Hawaii
for their warm hospitality. 
This research has made use of the NED (NASA extragalactic database) which is
operated by the Jet Propulsion Laboratory, California Institute of Technology,
under construct with the National Aeronautics and Space Administration.
TM is supported by a Research Fellowship from the Japan Society for the 
Promotion of Science for Young Scientists.
This work was financially supported in part by Grant-in-Aids for the Scientific
Research (Nos. 10044052, and 10304013) of the Japanese Ministry of Education,
Culture, Sports, and Science.

\clearpage
\end{document}